# Graphene on gallium arsenide: Engineering the visibility

M. Friedemann, K. Pierz, R. Stosch, and F. J. Ahlers

*Physikalisch-Technische Bundesanstalt,*

*Bundesallee 100, D-38116 Braunschweig, Germany*

Abstract:

Graphene consists of single or few layers of crystalline ordered carbon atoms. Its visibility on oxidized silicon (Si/SiO$_2$) enabled its discovery and spawned numerous studies of its unique electronic properties. The combination of graphene with the equally unique electronic material gallium arsenide (GaAs) has up to now lacked such easy visibility. Here we demonstrate that a deliberately tailored GaAs/AlAs (aluminum arsenide) multi-layer structure makes graphene just as visible on GaAs as on Si/SiO$_2$. We show that standard microscope images of exfoliated graphite on GaAs/AlAs suffice to identify mono-, bi-, and multi-layers of graphene. Raman data confirm our results.

Graphene is the most exciting material discovered[1,2] in the last decades, with a wealth of promising applications, from a 'QED-in-a-nutshell' system for fundamental studies[3], to ultra-fast electronics[4], use as an advanced molecular sensing material[5], and more[6]. Gallium arsenide (GaAs) is, on the other hand, the best understood semiconductor for ultra-fast electronics, opto-electronics and quantum-electronics applications[7]. Combining the two materials will certainly lead to additional opportunities, the use of the atomically flat surface of GaAs as the perfect host surface for graphene, or the probing of graphene with surface acoustic waves just being two immediate possibilities. However, a prerequisite for handling graphene, whether exfoliated from graphite[8] or grown elsewhere[9] and then transferred, is its easy visibility. Here we demonstrate how graphene is made visible on a specifically tailored GaAs/AlAs (aluminum arsenide) multi-layer substrate. Mono-, bi-, and multi-layers of graphene are identified by a simple contrast analysis of standard microscope images. The method is superior to Raman measurements which, however, are employed to verify that indeed graphene was prepared.

The fabrication of graphene devices used in many experimental studies currently relies on the fact that graphene crystallites can be visualized and identified using optical microscopy[10]. Silicon wafers with a top $SiO_2$ layer provide a good visibility even for monolayer graphene due to interference in the graphene/$SiO_2$/Si layer sequence. The high refractive index difference between 1.5 for $SiO_2$ and 4 for silicon and the transparency of $SiO_2$ lead to a significant change of the reflectivity $r_{gr}$ at positions where a graphene flake covers the surface. For certain oxide thicknesses the $SiO_2$/Si reflectivity $r_0$ has a broad minimum in the visible wavelength range, and at these wavelengths the graphene contrast $c := (r_0 - r_{gr})/r_0$ is largest[10,11]. This observation led us to the design rule which we applied to GaAs: Create a sequence of GaAs and AlAs layers which produces a low reflectivity in a not too narrow, and possibly visible, wavelength range. Such a layer sequence will make graphene visible, despite the much smaller refractive index difference from AlAs to GaAs (3 to 3.5) and the onset of AlAs absorption. We demonstrate this by the images in figure 1. The graphene was prepared with the same steps as for graphene on Si/SiO$_2$[8], using natural graphite as a starting material in an exfoliation process. The images were taken through a standard microscope objective of numerical aperture 0.9 (NA0.9) using a green filter with rectangular pass-band from 530 to 570 nm. Contrast values at the spots labeled M1 to F24 were extracted from these images (and from similar ones taken with NA0.5 and NA0.8 objectives) and are listed in Table I. As will be shown in detail below, the number of layers in a given graphene flake can be determined from these contrast data, and Raman measurements prove the results.

The design rule for a layer structure which leads to a sizeable graphene contrast was implemented as follows: for a given sequence of AlAs and GaAs layers the reflectivities $r_0$ of a graphene free surface and $r_{gr}$ of a graphene covered surface were calculated for a given angle of incidence and polarization by using Fresnel's law. When the total number of layers exceeds two or three it is convenient to use the transmission matrix formalism, described e. g. in Ref. [12]. For the real and imaginary parts of the refractive index we used tabulated data[13] for GaAs and AlAs, and data from Ref. [11] were used for Si and SiO$_2$, whose reflectivities were calculated for comparison. For graphene the complex refractive index given in Ref. [14] was used. We found that a layer stack of three double layers of AlAs on GaAs (each of thickness $d$) on top of a doped GaAs substrate was the simplest structure that met our design rule, while two double layers still had too high minimum reflectivity. The layers were then grown by molecular beam epitaxy (MBE) at a temperature above that normally employed for GaAs, in order to produce smooth AlAs interfaces, and they were capped with 5 nm of GaAs to protect the topmost AlAs layer from oxidation. In Figure 2a we compare their calculated reflectivity at normal incidence to that of silicon with 300 nm of SiO$_2$ (the most widely employed graphene host system), both for the case with graphene and without graphene. While in the viewgraph the influence of graphene is clearly visible for the silicon crystal where it causes a reduction of the reflectivity by nearly two percent absolute, it can hardly be distinguished on GaAs where it slightly increases the reflectivity. Yet the contrast $c = (r_0 - r_{gr})/r_0$ is of nearly the same size (Figure 2b, dashed lines) around 560 nm due to the designed low absolute reflectivity $r_0$ in this region. The wavelength of minimum reflectivity and hence maximum contrast is determined by the choice of the GaAs and AlAs layer thickness $d$, which was 36 nm in our case. For a calculation of the expected graphene contrast under realistic microscope imaging conditions one has to average the calculated reflectivity values over the range of angles of incident light determined by the objective's numerical aperture, and over the two polarization states. This calculation predicts a reduced contrast, compared to normal incidence. The result is presented by the solid lines in Figure 2b which were calculated for a numerical aperture of 0.8. The benefit that the loss of contrast at high apertures is less pronounced for GaAs than for silicon results from the term $n^2 - \sin^2(\theta)$, where the angle of incidence $\theta$ enters the Fresnel formulae: the large refractive index of AlAs is in this term much less affected by a large $\theta$ than the smaller refractive index of $SiO_2$.

As already mentioned, the GaAs and AlAs thickness $d$ is the parameter used to tune the contrast maximum to the desired wavelength. For a quantitative comparison of calculated contrasts with the measured values in Table I this thickness (or the wavelength, for that matter) must be known quite precisely since the optical filter pass-band and the high contrast band have comparable widths which could lead to errors when integrating over the filter pass-band. Therefore experimental reflectivity spectra of the MBE-grown AlAs/GaAs wafers were determined, and the thickness was obtained from these spectra by fitting $d$ to match the calculated to the measured reflectivity. We also took into account the effect of a surface oxide on GaAs whose typical thickness[15] is approximately 2 nm and which grows at the cost of the GaAs capping layer of the substrate. While the oxide layer itself has relatively little influence on the reflectivity, the ensuing reduction of the protective GaAs layer thickness reduces the distance between the wavelengths of maximum and minimum reflectivity. From this effect we estimated a GaAs cap layer thickness of 4 nm instead of the nominal value of 5 nm.

Using all this information the contrast values expected for microscope objectives with numerical apertures of 0.5, 0.8, and 0.9, and for a (550±20 nm) filter, were calculated and compared in Table I with the measured contrasts obtained from the images in Figure 1 (and similar ones for the other apertures). Already from this comparison it is clear that the flakes in Figure 1a consist of mono- and few-layer graphene. Not only do the calculated contrast values match the observed ones without using any fitting parameter, but also does our modeling describe the change of contrast with numerical aperture very well. With such a tool at hand the preparation of graphene on GaAs, and its purely microscope-optical identification for the post-processing steps become nearly as simple as on silicon.

As a final proof and in order to exclude any non-graphene related origin of the flakes in Figure 1, Raman measurements were performed at the spots M1 to F24, and GR. At the excitation wavelength of 532 nm the Raman lines were superposed by photo-luminescence likely originating from the AlAs layers, see Figure 3a. For the evaluation of the graphene Raman lines the broader spectral features from the AlAs luminescence were subtracted and corrected Raman spectra from the measurement spots M1 to T are shown in Figure 3b. As the contrast analysis already suggested the Raman 2D-line around 2675 cm$^{-1}$ (which is known to deliver a reliable 'fingerprint' identification of graphene[16]) is indeed from graphene. An unstructured 2D-line at the positions M1, M2, and M3 proves that graphene mono-layers exist at these spots. At spot B the 2D-line can be decomposed into four components, in good agreement with the findings for bi-layer graphene on silicon[16]. The tri-layer identification from Raman data is, however, less clear from the 2D line shape at spot T. That the graphene flake at T is different from that at B is, in the Raman spectra, better evident from their G- to 2D-line peak intensity ratios, which are visualized by bars in the inset of Figure 3b. Generally Raman spectra from graphitic material with three and more carbon mono-layers resemble more and more the spectrum of bulk graphite (see Figure 3a), and a determination of layer numbers from the Raman spectra alone becomes less reliable[17]. As we have shown above, the contrast analysis method is superior in this respect, except that it cannot deliver a material specific fingerprint proof.

In summary, we have shown how the optical properties of a substrate can be tailored to make exfoliated and transferred graphene layers visible. The method is so sensitive that it allows, when combined with a careful analysis of the materials' reflectivity, a determination of the number of graphene layers from the microscope images. We have used the method for the preparation of graphene on a GaAs surface. This material, the best studied semiconductor next to silicon, and due to its opto-electronic properties more versatile than silicon, will open numerous possibilities when combining its unique properties with those of graphene. The fact that in our study a very specific layer stack of AlAs and GaAs was used is no hindrance for a wider use of graphene on GaAs since the reflectivity response of other layer sequences, e. g. those hosting a high mobility two dimensional electron system, could be calculated and optimized to allow graphene identification on such structures.

This research has received funding from the European Community's Seventh Framework Programme, ERA-NET Plus, under Grant Agreement No. 217257. We acknowledge discussions with Kostya Novoselov, Stefano Borini, Uwe Siegner, and Egbert Buhr, as well as help with MBE growth by Holger Marx and with reflectivity measurements by Martina Kemlitz.

Table I

Contrast (in %) of graphene at spots M1 to F24 in Figure 1 determined from microscope images made with objectives of numerical aperture (NA) 0.5, 0.8, and 0.9. Calculated values are obtained as described in the text. In addition to specular reflectivity the calculation used a diffuse background reflectivity of 1.8% as determined from the difference of calculated to measured reflectivity. The number of layers used in the calculation is given in the last row.

|  |  | M1 | M2 | M3 | B | T | F8 | F13 | F24 |
|---|---|---|---|---|---|---|---|---|---|
| **NA0.5** | **meas.** | **7.0** | **7.0** | **6.9** | **10.3** | **16.5** | **50** | **93** | **175** |
|  | calc. | 4.8 | 4.8 | 4.8 | 9.9 | 15.1 | 43 | 74 | 143 |
| **NA0.8** | **meas.** | **5.5** | **5.4** | **5.4** | **9.1** | **14.0** |  |  |  |
|  | calc. | 4.2 | 4.2 | 4.2 | 8.6 | 13.2 |  |  |  |
| **NA0.9** | **meas.** | **3.9** | **3.8** | **3.5** | **7.4** | **10.7** | **30** | **53** | **113** |
|  | calc. | 3.8 | 3.8 | 3.8 | 7.8 | 12.0 | 35 | 62 | 125 |
| no of layers |  | 1 | 1 | 1 | 2 | 3 | 8 | 13 | 24 |

**Figure Captions**

Figure 1

Microscope images taken through a 550 ± 20 nm color filter and an objective with numerical aperture 0.9. The left image shows mainly mono-, bi-, and tri-layers of graphene, while in the right one comparatively thick graphite of eight and more mono-layers is shown. The contrast values at spots M1-F24 listed in Table I were determined from images like this. At the GR a Raman spectrum indistinguishable from that of bulk graphite was measured.

Figure 2

(color online) Comparison of calculated reflectivity and mono-layer graphene contrast for the GaAs/AlAs multi-layer described in the text (black (online blue) curves) and for silicon covered with 300 nm of $SiO_2$ (grey (online red) curves). (a) Reflectivity $r_0$ without graphene (solid line), and reflectivity $r_{gr}$ with graphene (dashed line). (b) Calculated contrast $c = (r_0 - r_{gr})/r_0$ for normally incident light (dashed line), and for light integrated over the aperture of a microscope objective with numerical aperture 0.8 (solid line). The GaAs contrast value was multiplied by -1 to allow easier comparison with the silicon case.

Figure 3

(a) Raman spectra (no vertical offset) from flakes M2 to GR. Raman lines G and 2D at 1587 cm$^{-1}$ and 2674 cm$^{-1}$ are accompanied by substrate luminescence. With increasing layer number $N$ (F8 to F24) the luminescence contributes less, and at spot GR the spectrum of graphite dominates, while the G-line shifts to $(1582.0 + 6.5 / N)$ cm$^{-1}$. (b) For the M1 to T spectra the background is subtracted, the horizontal scale is split, and the spectra are vertically offset. Unlike at B and T the G and the 2D line are unstructured at M1, M2, and M3. The 2D line from flake B is made up of four lines, characteristic of a graphene bi-layer Raman spectrum[16]. The data prove that the flakes M1, M2, and M3 are monolayers, while flake B is a bi-layer. For higher layer numbers the spectra resemble more that of graphite and the determination of the layer number from the 2D line shape is less clear[17], as indicated by the similar 2D-lines from flakes T and B. The relative peak intensity of the G to the 2D line (see inserted bar plot) is a more reliable indicator that flakes T and B are different. Already for $N = 3$ the contrast analysis is superior to the Raman method for determining the layer number.

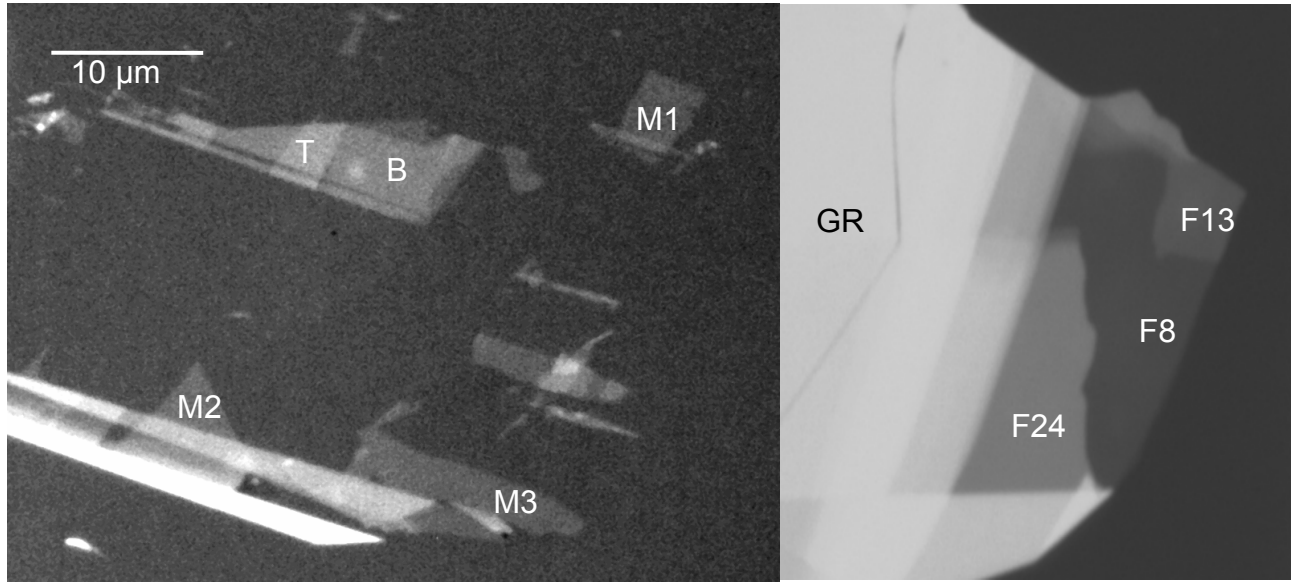

**Figure 1**

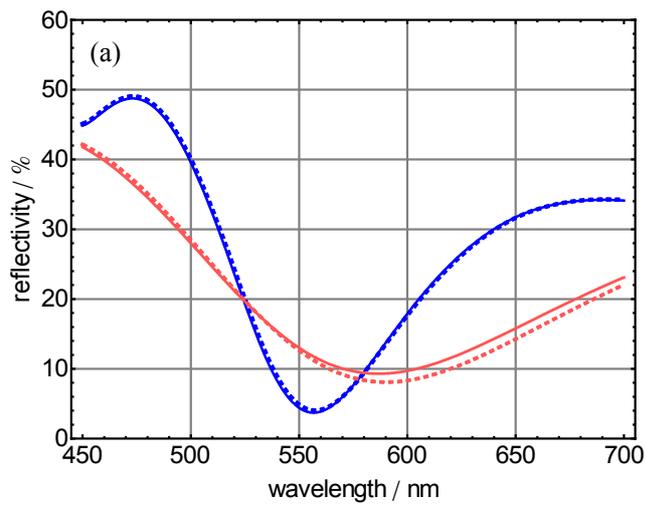 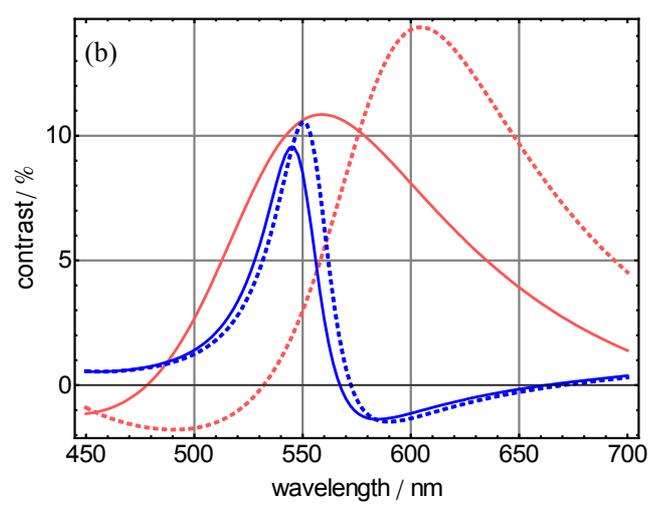

**Figure 2**

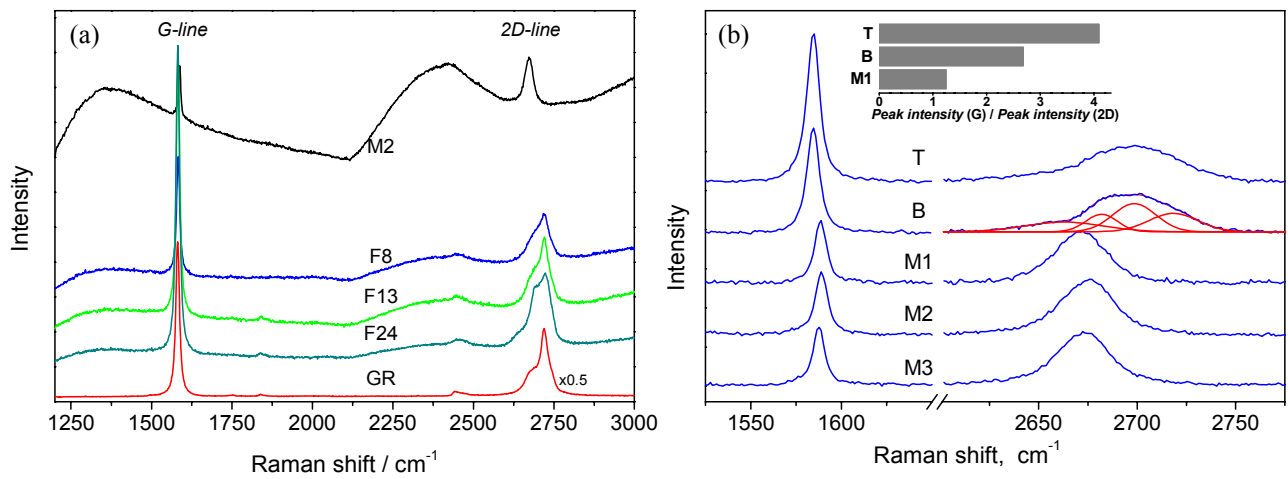

**Figure 3**